\documentclass[preprint,
superscriptaddress,
amsmath,amssymb,aps,showkeys,showpacs,
twoside,final,secnumarabic,%raggedbottom,
nofootinbib]{revtex4-2}

\usepackage[paperwidth=205mm,paperheight=290mm,top=17mm,bottom=25mm,
inner=17mm,outer=17mm,
twoside]{geometry}

\usepackage{cmap} % Улучшенный поиск %русских слов в~полученном pdf-файле
\usepackage[T1,T2A]{fontenc}
\usepackage[utf8]{inputenc}
\usepackage[russian,english]{babel}
\usepackage{color}
\usepackage{graphicx}% Include figure files
\usepackage{dcolumn}% Align table columns on decimal point
\usepackage{bm} % bold math
\usepackage[unicode=true,colorlinks=true,linkcolor=magenta, urlcolor=blue, citecolor = blue,breaklinks]{hyperref}
\usepackage{multirow}
\usepackage{url}
\usepackage{breakurl}
\DeclareGraphicsExtensions{.eps}

\newcount\issue
\newcount\Vol
\newcount\numb
\usepackage{fancyhdr} %this packages %provides fancy up and bottom of page
\def\Vol{\textbf{80}}
\def\numb{x}
\begin{document}

%====== Начало шапки статьи  ============
\title{RESONANCES AND STELLAR CYCLES:\\OBSERVATIONS AND  MODELLING\\[20pt]} 

\def\addressa{Sternberg Astronomical Institut, Lomonosov Moscow State University, Russia}
\def\addressb{Baku branch of Moscow State University, Azerbaijan}
\def\addressc{IZMIRAN, Troitsk, Russia}
\def\addressd{Faculty of Physics, Lomonosov Moscow State University, Russia}
\def\addresse{Space Research Institute, Moscow, Russia}

\author{\firstname{M.M.}~\surname{Katsova}}
\affiliation{\addressa}
\author{\firstname{F.A.}~\surname{Azizov}}
\affiliation{\addressb}
 \author{\firstname{V.N.}~\surname{Obridko}}
\affiliation{\addressc}
\author{\firstname{D.D.}~\surname{Sokoloff}}\email{E-mail: sokoloff.dd@gmail.com }
\affiliation{\addressc}
\affiliation{\addressd}
\author{\firstname{E.V.}~\surname{Yushkov}}
\affiliation{\addressd}
\affiliation{\addresse}

\received{xx.xx.2025}
\revised{xx.xx.2025}
\accepted{xx.xx.2025}

\begin{abstract}
In the paper we discuss the possibility of the influence of parametric excitation, in particular, planetary gravitational interaction, on the behavior of stellar magnetic activity cycles. Using the well-known Parker dynamo modeling, we demonstrate the doubtfulness of the fact that planetary rotation can be a determining factor in the formation of the cycle itself. However, we show that even a weak parametric influence can be sufficient to modulation of magnetic field oscillations, and, in particular, to the occurrence of beats. This result is discussed in the context of the influence of Jupiter on the occurrence of maxima and minima of the magnetic activity of our Sun.
\end{abstract}

%\pacs{Suggested PACS}\par
%DOI:  

\maketitle
\thispagestyle{fancy}

%====== Начало  статьи  ============

\section{Introduction}\label{intro}

The nominal 11-year solar activity cycle is a famous phenomenon important for the solar physics and many other branches of science. Similar activity cycles are known for some other late-type stars. The cycles are believed to be excited by solar/stellar dynamo driven by mirror-asymmetric solar/stellar convection. The point, however, is that Jupiter orbital period is about 11 years as well and it looks attractive at first sight to believe that a resonance between Jupiter orbital motion (planetary hypothesis) and solar magnetic field plays an important role here \cite{2019SolPh_Stephani}, \cite{Scaf22}. This idea as an alternative or addition to the dynamo explanation was suggested many times in various context in scientific literature. Of course, the idea deserves a comprehensive critical verification which, in particular, is performed by our team. The results of verification including some new findings are collected here. 

Just to be clear, our results taken in general look negative to the planetary hypothesis and in this way support the dynamo interpretation, however allows a possibility for some (limited) role of planetary effects in stellar dynamo.

\section{\label{sec:level1} Observational verification}
For quite a long time the magnetic activity periods and orbital frequencies in other stellar systems were unknown, so the point that the Jupiter case is just a coincidence looked like a probable statement which is impossible to support observationally. The observations, however, developed to allow desired verification, which was performed by \cite{Ketal1, Ketal2}. Contemporary astronomy knows several dozens stellar systems with known orbital periods of exoplanets and activity type of the host stars and never the orbital period coincides with the cycle length. Moreover, it looks that the availability of the planets does not inevitably yield in pronounced cyclic activity. The message is that planetary effects are, at least, not the leading drivers of stellar cycles.

%Let us move to the option that the planetary influence produces various deviations from the normal shape of the solar cycle, say, events like Maunder minimum. Tidal forces look like a relevant driver here. Of course, the tidal force contains the 11-year harmonic contribution from Jupiter orbital motion. Our point is that other planets give corresponding contributions as well, and celestial mechanics says that the resulting contribution is quite remote from a single harmonic oscillation. The solar activity temporal evolution is also quite remote from just an oscillation. From the viewpoint of the planetary hypothesis, one has to expect that both variations have to demonstrate some stable phase relation. Say, lower tidal forces results after dozen of years a lower cycle or something like that. Katsova et al. verified the expectation to learn that the actual track of the tidal force do not demonstrate a clear similarity to the solar activity evolution.

\section{Modulating solar dynamo drivers}
Consider the problem from the side of theory and ask ourselves how periodic modulation of dynamo drivers can modify the dynamo driven magnetic field evolution. The point here is that the dynamo equations are quite different from the pendulum harmonic equation, which is equal to a simple linear system of two ordinary differential equations, where a parametric periodic disturbance (in a certain frequency range) leads to a shift in the main frequency and to an exponential increase of oscillation amplitude. For a dynamo-system, the situation is slightly complex. In its simplest formulation (low-mode approximation) the dynamo-model can be reduced to a system of four ordinary differential equations. The asymptotic analysis of which, see \cite{2023JETP_Serenkova}, reveals the specific non-classical effects: suppression of the generation rate in the near-resonance region and frequency splitting instead of the shift.

To confirm the asymptotical results obtained for the low-mode approximation by \cite{2023JETP_Serenkova}, consider a numerical solution of the dynamo system in the simplest Parker statement. The Parker model is well known enough to be cited one more time, quite it should be mentioned that it describes the behavior of the toroidal and poloidal components of the magnetic field in a thin convective spherical layer, in particular, the typical solar cycle. Under the assumption of azimuthal symmetry, system consist of two partial (not ordinary) differential equations, whose behavior is determined by the differential rotation and hydromagnetic helicity. Here we use their product, so called, dynamo-number $D$, which parametric excitation is specified by a periodic harmonic ten percent change of this number with $\Omega$ frequency.

\begin{figure}[t]
\centering
\includegraphics[width = 0.47\textwidth]{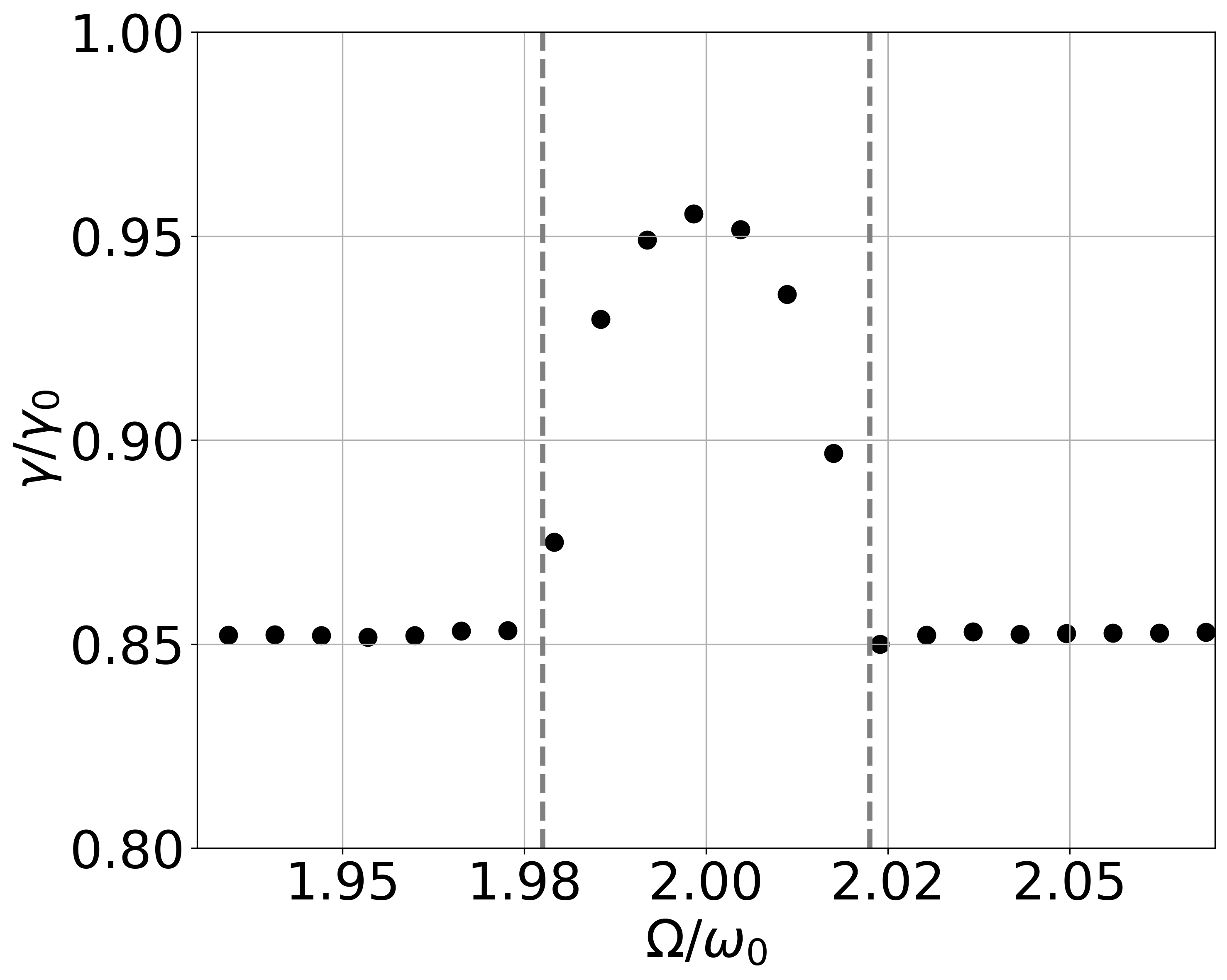}
\includegraphics[width = 0.47\textwidth]{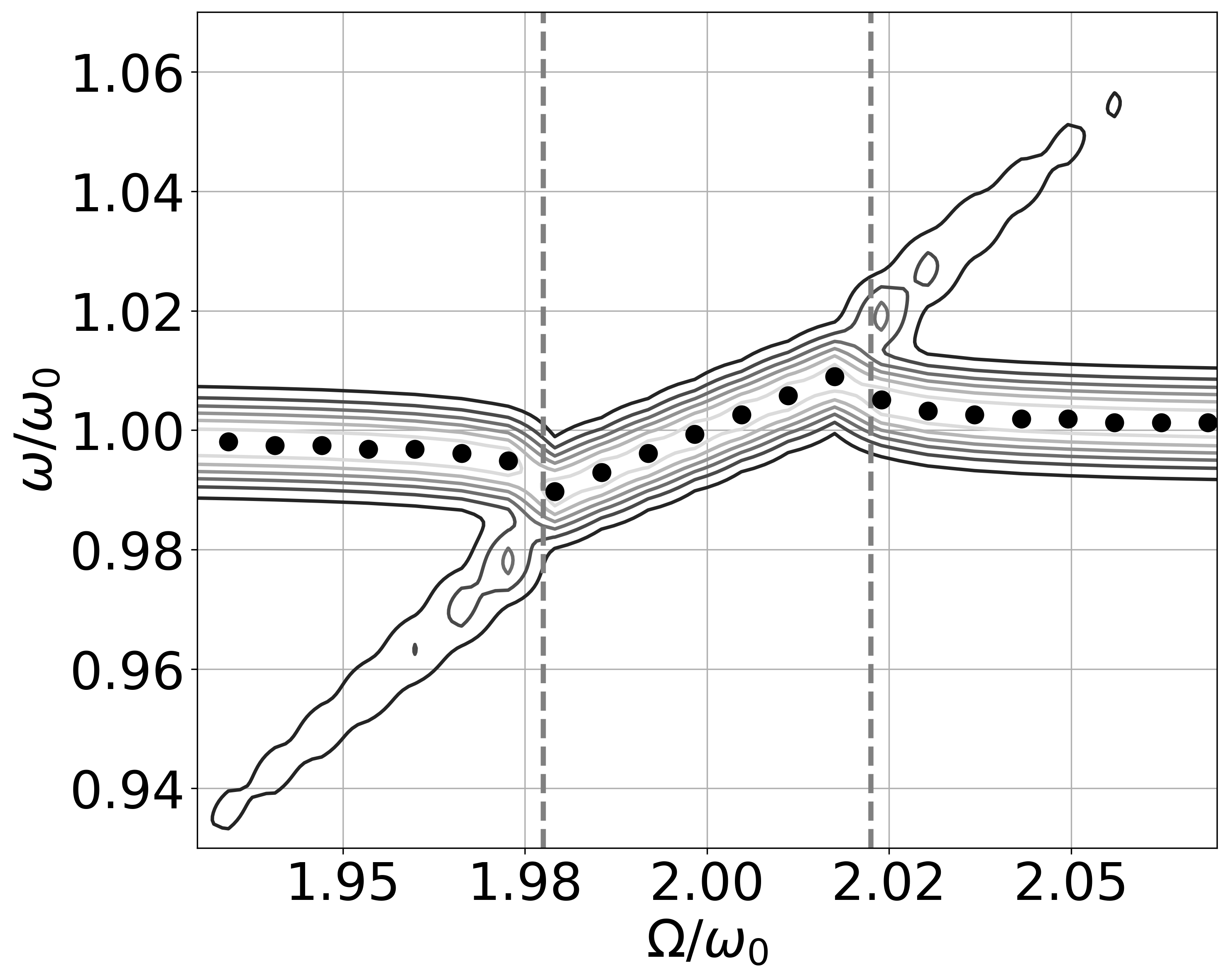}
\centering
\caption{Dependencies of the relative exponential growth rates $\gamma/\gamma_0$ (left panel) and the relative oscillation frequencies $\omega/\omega_0$ (right panel) on the relative parametric excitation frequency $\Omega/\omega_0$. Here we suppose ten percent excitation of the dynamo number: $D(t)=D_0(1+0.1\sin(\Omega t))$ and $D_0=750$.}
\label{Fig1}
\end{figure}

Calculating the exponential growth rates $\gamma$ and the oscillation frequencies $\omega$ of the Parker system solution relative to the parametric excitation frequencies $\Omega$, we obtain the dependencies shown on both panels the figure \ref{Fig1}. The left panel shows the dependence of the generation rate $\gamma/\gamma_0$, normalized to the growth rate of the magnetic field without parametric excitation $\gamma_0$, on the excitation frequency $\Omega/\omega_0$, normalized to the natural frequency of the system $\omega_0$. The resonance peak near doubled $\omega_0$ is quite clear, but during whole resonance region $\gamma/\gamma_0<1$, i.e. parametric excitation does not enhance, but suppresses the generation. Of course, if one consider a more effective excitation of the dynamo number -- not ten percent, but for example thirty percent, which was used in the recent work \cite{2025JETP_Farid} -- then at a double frequency the rate $\gamma/\gamma_0$ can be greater than one, but still, near the resonance region, growth rate suppression will be always observed.

Another distinctive feature is clearly visible in the right panel of the figure, which shows the spectral frequency decomposition of magnetic field oscillations depending on the normalized frequency of parametric excitation $\Omega/\omega_0$. The maximums of the spectral decomposition is marked with black circles and their dependency has a zigzag structure clearly demonstrating the frequency shift during parametric action. However, we can also see the splitting of the frequency near the resonance region. Such splitting leads to the beats of the toroidal and poloidal components, and the frequency of the beats is lower, the closer we are to the resonance region. It is clearly seen that a deviation from the double frequency by 2--3 percent is enough for the modulation of the system oscillations to appear. Moreover, the frequency range in which the beats are observed is greater than the range in which the resonance peak is located.

\section{Discussion and conclusions}
Both observational data and numerical modeling data clearly show that planetary parametric excitation is unlikely to be the cause of stellar activity cycles. Rather, the parametric effect may lead not to an increase, but to a suppression of the generation rate. This suppression is especially evident in a wide frequency range in the near-resonance region. On the other hand, numerical and asymptotic analysis of the simplest dynamo models, in particular, the low-mode or Parker model, show that parametric action leads to a non-classical effect of frequency splitting and the appearance of oscillation modulation. This effect is unexpected, since it is absent for parametric resonance in the classical equation of harmonic oscillations. On the other hand, it is very probable in the context of changing stellar activity cycles and the formation of local minima and maxima.

Here it should be emphasized once again that the real picture of the modulation of stellar cycles can be much more complex than the formation of beats. Since it is strange to expect that all the nuances of the dynamics of magnetic activity, for example, of the Sun, will be taken into account in the simplest model considered. However, even a three-five percent difference in the proper period of solar activity and the orbital period  of Jupiter, according to the data obtained, may well lead to a modulation with a characteristic period of 100--200 years.

The study of M.M.Katsova  was conducted under the state assignment of Lomonosov Moscow State University. 
The work of E.V.Yushkov on the asymptotic analysis of the parametric resonance was supported by the grant of the Russian Science Foundation number 23-11-000 69. 

Data Availability: the data on sunspots were taken from WDC-SILSO, Royal Observatory of Belgium, Brussels,
\url{https://sidc.be/SILSO/datafiles}.

%%%%%%%%%%%%%%%%%%%%%%%%%%%%%%%%
% USE thebibliography
%%%%%%%%%%%%%%%%%%%%%%%%%%%%%%%%

\end{document}